\documentclass[aps,preprint,prd,showpacs,nofootinbib]{revtex4-1}
\usepackage{graphicx}
\usepackage{bm}
\usepackage{times}
\usepackage{hyperref}
\usepackage{slashed}
\usepackage{amsmath}
\usepackage{xcolor}
\usepackage{fancyhdr}
\usepackage{footmisc}
\usepackage{soul}
\usepackage{lipsum}
\usepackage{multirow}
\pdfoutput=1
\usepackage{amssymb}
\usepackage{hyperref}
\newcommand{\be}{\begin{equation}}
\newcommand{\ee}{\end{equation}}
\newcommand{\bea}{\begin{eqnarray}}
\newcommand{\eea}{\end{eqnarray}}
\newcommand{\bes}{\begin{subequations}}
\newcommand{\ees}{\end{subequations}}

\newcommand{\bc}{\begin{center}}
\newcommand{\ec}{\end{center}}
\usepackage{amsmath, amssymb} 
\usepackage{slashed}
\usepackage{natbib}

\begin{document}
\title{ Flavor Changing Neutral Current processes and family discrimination in  3-3-1 models }
\author{Vin\'icius Oliveira$^a$}\email{vlbo@academico.ufpb.br
}  
\author{C. A. de S. Pires$^{a}$}\email{cpires@fisica.ufpb.br} 
\affiliation{$^{a}$Departamento de F\'isica, Universidade Federal da Para\'iba, Caixa Postal 5008, 58051-970, Jo\~ao Pessoa, PB, Brazil}

\date{\today}
\vspace{1cm}
\begin{abstract}
In 3-3-1 models anomaly cancellation requires that one of the three families of quarks transforms as triplet by $\text{SU}(3)_\text{L}$   with the other two transforming necessarily as anti-triplet. This is an important feature of the  model because with it we explain family replication. Thus it is mandatory to discriminate which of the families will transform as triplet by $\text{SU}(3)_\text{L}$  because the main consequence of anomaly cancellation in 3-3-1 models is the  arising of processes  violating flavor at tree level by means of neutral currents mediated by gauge and scalar fields and each case leads to different results. In this work  we consider the 3-3-1 model with right-handed neutrinos. Among the spectrum of 3-3-1 particles that contributes to the flavor changing neutral processes, there is a pseudoscalar that may be the lightest of the 3-3-1 particles and then should  give the main contribution to such processes. We then calculate its contribution  to the $K^0 -\bar K^0$ mixing transition and confront it with the current experimental results. We do this for the three cases in which one of the  family of quarks transforms as a triplet by $\text{SU}(3)_\text{L}$. According to our findings each case leads to different constraints on the mass of the pseudoscalar and  the case in which the third family of quarks transforms as triplet seems to be the favoured one. We also obtain the most stringent bounds on the mass of the pseudoscalar of the model. 
\end{abstract}

\maketitle
\section{Introduction}

The 3-3-1 model with right-handed neutrinos \cite{Singer:1980sw, Foot:1994ym,Montero:1992jk} (331RHN for short) is a phenomenologically  viable version of models based on the $\text{SU}(3)_\text{C} \times \text{SU}(3)_\text{L} \times \text{U}(1)_\text{N}$ gauge symmetry\footnote{For other versions, see: \cite{Frampton:1992wt,Pisano:1992bxx,Ozer:1995xi,Sanchez:2001ua}.}. As main  theoretical aspects of this class of models we have that they explain electric charge quantization\cite{deSousaPires:1999ca,deSousaPires:1998jc}, provide a natural solution to the strong CP-problem\cite{Pal:1994ba}, may be unified into a $SU(6)$ model\cite{Li:2019qxy}, but its remarkable theoretical aspect is that this class of models requires exactly three families of fermions to have theoretical consistency. In other words, the model  explains, in this way, family replication up to the third family. This is so because  anomaly is not canceled within each generation \cite{Pisano:1996ht,Frampton:1992wt}. It is canceled among three generations \cite{Ng:1992st,Liu:1993gy,Pisano:1996ht}.\footnote{The anomalies in question are: $[SU(3)]^2U(1)_N$, $[SU(3)_C]^2U(1)_N$ and $[U(1)_N]^3$.} Then fermions must come in equal numbers of triplet and anti-triplet representations of $\text{SU}(3)_\text{L}$. Leptons are treated democratically with the three families coming in triplet representation. This avoids Flavor Changing Neutral Current (FCNC) among the leptons. Consequently two families of quarks must come in anti-triplet representation. Such arrangement of quark families necessarily generates FCNC processes among the quarks. In fact such processes are the main signature of the model and receive contributions from all neutral scalars and gauge bosons of the model, except the photon.

The original scalar content of the 331RHN involves three triplets of scalars and after spontaneous breaking of the 3-3-1 symmetry the spectrum of scalars recovers what we call an effective 2HDM plus three singlet of scalars whose masses belong to the 3-3-1 scale \cite{Pinheiro:2022bcs,Cherchiglia:2022zfy,Ng:1992st}. Current constraints impose that the particle of the effective 2HDM must have mass of hundreds of GeVs. For example, the pseudoscalar can not be lighter than $350$ GeV \cite{Cherchiglia:2022zfy}. In what concerns the gauge sector, the model contains nine gauge bosons, namely the standard gauge bosons plus $V^{\pm}$, $U^0$, $U^{0 \dagger}$ and $Z^{\prime}$. The structure of the gauge bosons are the same one of the standard gauge bosons with the difference that in the neutral sector we are going to have FCNC \cite{Long:1996rfd}. LHC current constraint imposes $Z^{\prime}$ mass around $4$ TeV \cite{Coutinho:2013lta, Alves:2022hcp} what implies that the gauge bosons $V^{\pm}$ and $U^0$ have masses around this value, too \cite{Long:1996rfd}.

Theory is not able to discriminate which family of quark must come in triplet representation. Previous studies in this direction go back  to the nineties. The first  study was done in the minimal 3-3-1 model \cite{Frampton:1992wt,Pisano:1992bxx}  and concluded that the third family should transform as a triplet \cite{Ng:1992st}. The same study was done in the 331RHN and also concluded that the third family should be treated differently from the other two \cite{Long:1999ij}. Both studies considered mesons transitions with  $Z^{\prime}$ giving the dominant contribution. Although flavor physics has been intensively studied in the context of the  model \cite{Buras:2012dp,Buras:2014yna,Buras:2016dxz,Buras:2023fhi,Benavides:2009cn,Rodriguez:2004mw, CarcamoHernandez:2022fvl}, as far as we know the studies done in Refs. \cite{Ng:1992st,Long:1999ij} are the unique works using FCNC to discriminate families in 3-3-1  models . 

Here we attack the problem of family discrimination  but now assuming that it is the pseudoscalar that gives the main contributions to such processes. This is justified by the studies done in Refs. \cite{Pinheiro:2022bcs, Cherchiglia:2022zfy} that indicates that the pseudoscalar may be the lightest of the 3-3-1 particles that contributes to such processes \cite{Long:1999ij,Ng:1992st,deSPires:2007wat}. We then consider the $K^0 - \bar K^0$ transition and investigate if the contribution of such pseudoscalar  is able to give us any clue about family discrimination.  We do this for the three possibilities (variants) of arrangement of families.  We will see that  each case leads to different constraints on the mass of the pseudoscalar allowing, in this way, to  discriminate which variant is favoured by the model. Moreover, we obtain the most stringent bound on the mass of such pseudoscalar.

This paper is organized in the following way: in \autoref{sec:2} we summarize the essence of the model; in \autoref{sec:3} we discuss the possibles choices for quarks anti-triplet representation; in \autoref{sec:4} we derive the mass difference terms for the neutral mesons; in \autoref{sec:5} we present the numerical results; finally we summarize and draw our conclusions in \autoref{sec:6}.

\section{The main aspects of the model} \label{sec:2}

In the 331RHN right-handed neutrinos compose the third component of the leptonic triplets, $f_l=(\nu_{l_L}\,,\, e_{l_L} \,,\, \nu^C_{l_R})^T$ \cite{Foot:1994ym,Montero:1992jk}, where $l=e\,\,,\ \mu \,,\, \tau$. The gauge sector of the model involves nine gauge bosons related to the electroweak sector \cite{Long:1996rfd} where four of them are the standard gauge bosons $W^{\pm}$, $ Z^0$,  $\gamma$, the other  five  $W^{\prime \pm}$,  $U^0$, $U^{0 \dagger}$, and $Z^{\prime}$ are the typical 3-3-1 gauge bosons, and eight gluons $g$. We describe the quark sector in the next section.

The original scalar sector of the  model is composed by three triplets of scalars
\begin{eqnarray}
\eta = \left (
\begin{array}{c}
\eta^0 \\
\eta^- \\
\eta^{\prime 0}
\end{array}
\right ),\,\rho = \left (
\begin{array}{c}
\rho^+ \\
\rho^0 \\
\rho^{\prime +}
\end{array}
\right ),\,
\chi = \left (
\begin{array}{c}
\chi^0 \\
\chi^{-} \\
\chi^{\prime 0}
\end{array}
\right ),
\label{scalarcont} 
\end{eqnarray}
with $\eta$ and $\chi$ transforming as $(1\,,\,3\,,\,-1/3)$
and $\rho$ as $(1\,,\,3\,,\,2/3)$.

The most economical potential of the model is obtained by demanding  lepton number conservation at tree level \cite{Pal:1994ba,deSPires:2007wat}. In this case we have:
\begin{eqnarray} 
V(\eta,\rho,\chi)&=&\mu_\chi^2 \chi^2 +\mu_\eta^2\eta^2
+\mu_\rho^2\rho^2+\lambda_1\chi^4 +\lambda_2\eta^4
+\lambda_3\rho^4+ \nonumber \\
&&\lambda_4(\chi^{\dagger}\chi)(\eta^{\dagger}\eta)
+\lambda_5(\chi^{\dagger}\chi)(\rho^{\dagger}\rho)+\lambda_6
(\eta^{\dagger}\eta)(\rho^{\dagger}\rho)+ \nonumber \\
&&\lambda_7(\chi^{\dagger}\eta)(\eta^{\dagger}\chi)
+\lambda_8(\chi^{\dagger}\rho)(\rho^{\dagger}\chi)+\lambda_9
(\eta^{\dagger}\rho)(\rho^{\dagger}\eta) \nonumber \\
&&-\frac{f}{\sqrt{2}}\epsilon^{ijk}\eta_i \rho_j \chi_k +\mbox{H.c.}\,.
\label{potentialII}
\end{eqnarray}

We restrict our approach to the case where only $\eta^0\,\,, \rho^0$, and $\chi^{\prime 0}$ develop VEV. We then shift the fields  in the usual way
\begin{equation}
  \eta^0\,\,, \rho^0\,\,, \chi^{\prime 0} =\frac{1}{\sqrt{2}}(v_{\eta\,, \rho\,,\chi^{\prime}}+R_{_{\eta\,, \rho\,,\chi^{\prime}}}+iI_{_{\eta\,, \rho\,,\chi^{\prime}}})\,.
\end{equation}
The potential above provides the following set of  minimal constraint equations 
\begin{eqnarray}
 &&\mu^2_\chi +\lambda_1 v^2_{\chi^{\prime}} +
\frac{\lambda_4}{2}v^2_\eta  +
\frac{\lambda_5}{2}v^2_\rho-\frac{f}{2}\frac{v_\eta v_\rho}
{ v_{\chi^{\prime}}}=0,\nonumber \\
&&\mu^2_\eta +\lambda_2 v^2_\eta +
\frac{\lambda_4}{2} v^2_{\chi^{\prime}}
 +\frac{\lambda_6}{2}v^2_\rho -\frac{f}{2}\frac{v_{\chi^{\prime}} v_\rho}
{ v_\eta} =0,
\nonumber \\
&&
\mu^2_\rho +\lambda_3 v^2_\rho + \frac{\lambda_5}{2}
v^2_{\chi^{\prime}} +\frac{\lambda_6}{2}
v^2_\eta-\frac{f}{2}\frac{v_\eta v_{\chi^{\prime}}}{v_\rho} =0.
\label{mincond} 
\end{eqnarray}

After spontaneous breaking of the 3-3-1  symmetry  the scalar sector get composed by three CP-even scalars, $h$, $H_1$, and $H_2$ where the first recovers the features of the standard Higgs and the other two are heavy scalars with mass proportional to $v_{\chi^{\prime}}$\footnote{For the development of the scalar sector of this model , see Refs. \cite{deSPires:2007wat, Pinheiro:2022bcs}. These three neutral CP-even scalars contribute to FCNC processes. We assume the $A$ is lighter than $H_1$and $H_2$. Then we neglect the contributions of $H_{1,2}$ to FCNC processes.}. The complex scalar $\chi^0$ is a Goldstone eaten by the non-hermitian neutral gauge boson $U^0$ and $\eta^{\prime 0}$ decouple from the other scalars and do not contribute to FCNC. There are also two heavy singly charged scalars\footnote{The contribution of these scalars to the process $b\rightarrow s \gamma$ imposes a lower bound of $650$ GeV to the lightest charged scalar. } and one pseudoscalar $A$. Our interest here lies in $A$. The justification comes below.

Considering the basis $(I_{\chi^{\prime}}, I_\eta, I_\rho)$, the mass matrix that involves the CP-odd scalars are
\begin{equation}
M_I^2=
\begin{pmatrix}
fv_\eta v_\rho/4v_{\chi^{\prime}} &  f v_\rho/4 &  f v_\eta/4\\
f v_\rho/4 &  fv_{\chi^{\prime}} v_\rho/4v_\eta & f v_{\chi^{\prime}}/4\\
 f v_\eta/4 &  f v_{\chi^{\prime}}/4 &  fv_{\chi^{\prime}} v_\eta/4v_\rho
\end{pmatrix}.
\label{MI}
\end{equation}

Besides the complexity of $M_I^2$, its texture presents a high level of symmetry such that its diagonalization leads to  two null eigenvalues. In order to simplify the level of complexity, we also assume throughout this work that  $v_{\chi^{\prime}} \gg v_\eta , v_\rho$ and that  $v_\eta=v_\rho=v$. In this case the  eigenvectors   $P \sim I_{\chi^{\prime}}$ and $G \approx  \frac{1}{\sqrt{2}} I_\eta -\frac{1}{\sqrt{2}} I_\rho$ represent the goldstone bosons eaten by the gauge bosons $Z^{\prime}$ and $Z$ \cite{Cao:2016uur,deSPires:2007wat}. The third eigenvector,  $A= \frac{1}{\sqrt{2}} I_\eta +\frac{1}{\sqrt{2}} I_\rho$,   has mass given by  $m^2 _A=\frac{f}{2}v_{\chi^{\prime}}$. Observe that $m_A$ depends directly on $f$ which is a free parameter of the model. The LHC lower bound on $v_{\chi^{\prime}}$ found in Refs.  \cite{Coutinho:2013lta, Alves:2022hcp} requires $v_{\chi^{\prime}}>10^4$ GeV. Then, whatever the value of $v_{\chi^{\prime}}$ is, the parameter $f$ may  lower or rise $m_A$. As there are no constraints on $f$ it then may acquire values such that make $A$ the lightest of the 3-3-1 particle that mediates FCNC processes as $K^0 -\bar K^0$ mixing. Here we assume that  this is the case, namely  $A$ is supposed to give the dominant contribution to $\Delta m_K$\footnote{For previous studies of FCNC in 3-3-1 models, see: \cite{Ng:1992st,Liu:1993gy,GomezDumm:1993oxo,Long:1999ij,Rodriguez:2004mw,Cabarcas:2007my,Promberger:2007py,Benavides:2009cn,Cabarcas:2009vb,Cabarcas:2011hb,Jaramillo:2011qu,Cogollo:2012ek,Machado:2013jca,Okada:2016whh,Queiroz:2016gif,Huitu:2019kbm,NguyenTuan:2020xls}.}.  This is the reason why we focus on $A$ to investigate the role of family discrimination in FCNC processes.   In what follows we obtain the Yukawa interactions composed by  $A$ and the quarks that contribute to  meson mixing transitions for the three variants of interest here\footnote{ We stress that leptons gain masses by means of the Yukawa interactions $Y_{ll}\bar f_{l_L} \rho e_{l_R}$ and as the three families of leptons transform as triplet by $SU(3)_L$, then the leptons did not get involved in  FCNC processes \cite{Long:1996rfd}.}.

\section{Variants}\label{sec:3}

\subsection{Variant I}
In this case the first two families of quarks  transform as anti-triplet while the third one  transforms as triplet by $\text{SU}(3)_\text{L}$
\begin{eqnarray}
&&Q_{i_L} = \left (
\begin{array}{c}
d_{i} \\
-u_{i} \\
d^{\prime}_{i}
\end{array}
\right )_L\sim(3\,,\,\bar{3}\,,\,0)\,,u_{iR}\,\sim(3,1,2/3),\,\,\,\nonumber \\
&&\,\,d_{iR}\,\sim(3,1,-1/3)\,,\,\,\,\, d^{\prime}_{iR}\,\sim(3,1,-1/3),\nonumber \\
&&Q_{3L} = \left (
\begin{array}{c}
u_{3} \\
d_{3} \\
u^{\prime}_{3}
\end{array}
\right )_L\sim(3\,,\,3\,,\,1/3),u_{3R}\,\sim(3,1,2/3),\nonumber \\
&&\,\,d_{3R}\,\sim(3,1,-1/3)\,,\,u^{\prime}_{3R}\,\sim(3,1,2/3),
\label{quarks} 
\end{eqnarray}
where  the index $i=1,2$ is restricted to only two generations. The negative signal in the anti-triplet $Q_{i_L}$ is just to standardise the signals of the charged current interactions with the gauge bosons.  The primed quarks are new heavy quarks with the usual $(+\frac{2}{3}, -\frac{1}{3})$ electric charges. 

Here the simplest  Yukawa interactions that generate the correct mass for all standard quarks are composed by the terms\footnote{For the most general Yukawa interactions involving terms that violate lepton number, see: \cite{Doff:2006rt}.},
\begin{equation}\label{yukawa}
-{\cal L}_Y \supset g^1_{ia} \bar Q_{i_L} \eta^* d_{a_R} + h^1_{3a} \bar Q_{3_L} \eta u_{a_R} + g^1_{3a} \bar Q_{3_L} \rho d_{a_R} + h^1_{ia} \bar Q_{i_L} \rho^* u_{a_R} + \mbox{H.c.}\,,
\end{equation}
where $a=1,2,3$ and the parameters $g^1_{ab}$ and $h^1_{ab}$ are Yukawa couplings that, for sake of simplification,  we consider reals.

Let us consider the standard up quarks. For the basis $u=(u_1\,,\,u_2\,,\, u_3)$  we get the following mass matrix
\begin{equation}
M_u=\frac{v}{\sqrt{2}}
\begin{pmatrix}
-h^1_{11} & -h^1_{12} & -h^1_{13} \\
-h^1_{21} & -h^1_{22}  & -h^1_{23}\\
h^1_{31} & h^1_{32} & h^1_{33}
\end{pmatrix},
\label{Mu}
\end{equation}
where the negative signals in this matrix arise due to the negative signal in $Q_{i_L}$ above.  Diagonalizing this matrix by a bi-unitary transformation
\begin{equation}
V_L^{u \dagger}M_u V_R^u=
\begin{pmatrix}
m_u & 0 & 0 \\
0 & m_c  & 0\\
0 & 0 & m_t
\end{pmatrix},
\label{MuD}
\end{equation}
we get the masses of the up quarks.  The relation among the basis is given by
\begin{equation}
   \hat u_{L,R}=V^{\dagger u}_{L,R} u_{L,R}\,,
\end{equation}
with $\hat u=(u\,\,,\,\,c\,\,,\,\, t)^T$.

For the down quarks we  have the mass matrix
\begin{equation}
M_d=\frac{v}{\sqrt{2}}
\begin{pmatrix}
g^1_{11} & g^1_{12} & g^1_{13} \\
g^1_{21} & g^1_{22}  & g^1_{23}\\
g^1_{31} & g^1_{32} & g^1_{33}
\end{pmatrix},
\label{Md}
\end{equation}
diagonalizing this matrix by a bi-unitary transformation
\begin{equation}
V_L^{d \dagger}M_d V_R^d=
\begin{pmatrix}
m_d & 0 & 0 \\
0 & m_s  & 0\\
0 & 0 & m_b
\end{pmatrix},
\label{MdD}
\end{equation}
we get the masses of the down quarks. The relation among the basis is given by
\begin{equation}
   \hat d_{L,R}=V^{\dagger d}_{L,R} d_{L,R}
\end{equation}
with $\hat d=(d\,\,,\,\,s\,\,,\,\, b)^T$. The mixing matrices $V_{L}^{d,u}$ are the most general $3\times 3$ real matrix that obey two constraint, namely they must be unitary, $V_L^{u \dagger} V^u_L=I\,,\,\,\,V^{d \dagger}_L V^d_L=I$, and obey the constraint  $V^{u }_L V^{d \dagger}_L=V_\text{CKM}$, where \cite{Workman:2022ynf}
\begin{equation}
|V_\text{CKM}|=
\begin{pmatrix}
0.97435  & 0.22500  & 0.00369 \\
0.22486  & 0.97349   & 0.04182\\
0.00857 & 0.04110 & 0.999118
\end{pmatrix},
\label{CKMexp}
\end{equation}
is the CKM matrix.

Note that $V_{R}^{d,u}$ play no role at all. The usual assumption here is to take $V_{R}^{d,u}=I$. In this case the relations in  Eqs. (\ref{Mu}), (\ref{MuD}), (\ref{Md}) and (\ref{MdD}) allow we write the Yukawa coupling as 
\begin{eqnarray}
   && g^1_{ia}=\sqrt{2}(V^d_L)_{ia}\frac{(m_{down})_a}{v}\,, \, g^1_{3a}=\sqrt{2}(V^d_L)_{3a}\frac{(m_{down})_a}{v}\,,\nonumber \\
   &&  h^1_{ia}=-\sqrt{2}(V^u_L)_{ia}\frac{(m_{up})_a}{v}\,, \, h^1_{3a}=\sqrt{2}(V^u_L)_{3a}\frac{(m_{up})_a}{v},
\end{eqnarray}
where $i=1,2$, $a=1,2,3$,  $(m_{down})_a=m_d, m_s,m_b$, and  $(m_{up})_a=m_u, m_c,m_t$.

After all this,  in the end of the day we obtain the following Yukawa interactions among $A$ and the physical standard quarks
\begin{eqnarray}
 {\cal L}_Y^{A}&=&iA  \bar{\hat{u}}_{b_L}\left( \frac{1}{\sqrt{2}} (V^u_L)_{3a} (V^u_L)_{b3}\frac{(m_{up})_a}{v}   -\frac{1}{\sqrt{2}} (V^u_L)_{ia} (V^u_L)_{bi}\frac{(m_{up})_a}{v} \right)\hat u_{a_R} \nonumber \\
&+& iA \bar{ \hat{d}}_{b_L}\left( -\frac{1}{\sqrt{2}} (V^d_L)_{ia} (V^d_L)_{bi}\frac{(m_{dowm})_a}{v}   +\frac{1}{\sqrt{2}} (V^d_L)_{3a} (V^d_L)_{b3}\frac{(m_{dowm})_a}{v}\right)\hat d_{a_R}  + \text{H.c.}\,,
\label{YcaseI}
\end{eqnarray}
with the subscript $b=1,2,3$. Observe that the interactions above that lead to FCNC processes  depend on the elements of the mixing matrices $(V^{u,d}_{L})_{ab}$ which are free parameters. The common procedure here is to use FCNC processes to constrain the mass of the particle that  mediates the process, in our case the pseudo-scalar $A$. Then people postulate a specific texture for $V^{u,d}_L$. We follow this approach here with a remarkable improvement. We return to this point in Sec. V.
\subsection{Variant II}
In this case the first family transforms as triplet and the second and third family transform as anti-triplet, which means
\begin{eqnarray}
&&Q_{1_L} = \left (
\begin{array}{c}
u_{1} \\
d_{1} \\
u^{\prime}_{1}
\end{array}
\right )_L\sim(3\,,\,3\,,\,1/3)\,,u_{1R}\,\sim(3,1,2/3),\,\,\,\nonumber \\
&&\,\,d_{1R}\,\sim(3,1,-1/3)\,,\,\,\,\, u^{\prime}_{iR}\,\sim(3,1,2/3),\nonumber \\
&&Q_{iL} = \left (
\begin{array}{c}
d_{i} \\
-u_{i} \\
d^{\prime}_{i}
\end{array}
\right )_L\sim(3\,,\,\bar 3\,,\,0),u_{3R}\,\sim(3,1,2/3),\nonumber \\
&&\,\,d_{3R}\,\sim(3,1,-1/3)\,,\,d^{\prime}_{3R}\,\sim(3,1,-1/3),
\label{quarks} 
\end{eqnarray}
where  the index $i=2,3$ is restricted to only two generations. 

The minimal set of Yukawa interactions that leads to  the correct quark masses involves the terms,
\begin{eqnarray}
&-&{\cal L}_Y \supset  g^2_{1a}\bar Q_{1_L}\rho d_{a_R}+ g^2_{ia}\bar Q_{i_L}\eta^* d_{a_R} \nonumber \\
&&+h^2_{1a} \bar Q_{1_L}\eta u_{a_R} +h^2_{ia}\bar Q_{i_L}\rho^* u_{a_R} + \mbox{H.c.}\,.
\label{yukawa1}
\end{eqnarray}
Following all the previous procedures done in Variant I, we obtain the following Yukawa interactions among $A$ and the standard quarks
\begin{eqnarray}
 {\cal L}_Y^{A}&=& iA \bar{\hat{u}}_{b_L} \left( \frac{1}{\sqrt{2}} (V^u_L)_{1a} (V^u_L)_{b1}\frac{(m_{up})_a}{v}   -\frac{1}{\sqrt{2}} (V^u_L)_{ia} (V^u_L)_{bi}\frac{(m_{up})_a}{v} \right)\hat u_{a_R}  \nonumber \\
&+& iA\bar{ \hat{d}}_{b_L}\left( \frac{1}{\sqrt{2}} (V^d_L)_{1a} (V^d_L)_{b1}\frac{(m_{dowm})_a}{v}    -\frac{1}{\sqrt{2}} (V^d_L)_{ia} (V^d_L)_{bi}\frac{(m_{dowm})_a}{v}\right)\hat d_{a_R}  + \mbox{H.c.}\,,
\label{YcaseII}
\end{eqnarray}
with the subscript $b=1,2,3$.

\subsection{Variant III}

In this case the second family transforms as triplet while the first and third transform as anti-triplet
\begin{eqnarray}
&&Q_{2_L} = \left (
\begin{array}{c}
u_{2} \\
d_{2} \\
u^{\prime}_{2}
\end{array}
\right )_L\sim(3\,,\,3\,,\,1/3)\,,u_{2R}\,\sim(3,1,2/3),\,\,\,\nonumber \\
&&\,\,d_{2R}\,\sim(3,1,-1/3)\,,\,\,\,\, u^{\prime}_{2R}\,\sim(3,1,2/3),\nonumber \\
&&Q_{iL} = \left (
\begin{array}{c}
d_{i} \\
-u_{i} \\
d^{\prime}_{i}
\end{array}
\right )_L\sim(3\,,\,\bar 3\,,\,0),u_{iR}\,\sim(3,1,2/3),\nonumber \\
&&\,\,d_{iR}\,\sim(3,1,-1/3)\,,\,d^{\prime}_{iR}\,\sim(3,1,-1/3),
\label{quarks} 
\end{eqnarray}
where $i=1,3$.

Here the Yukawa interactions among the scalars and the standard  quarks are
\begin{eqnarray}
&-&{\cal L}_Y \supset g^3_{2a}\bar Q_{2_L}\rho d_{a_R}+ g^3_{ia}\bar Q_{i_L}\eta^* d_{a_R} \nonumber \\
&&+h^3_{2a} \bar Q_{2_L}\eta u_{a_R} +h^3_{ia}\bar Q_{i_L}\rho^* u_{a_R} + \mbox{H.c.},
\label{yukawa1}
\end{eqnarray}
where $i=1,3$.

Following the procedure of Variant I the interactions of the pseudoscalar $A$ with the standard quarks are
\begin{eqnarray}
 {\cal L}_Y^{A}&=& iA \bar{\hat{d}}_{b_L} \left( \frac{1}{\sqrt{2}} (V^d_L)_{2a} (V^d_L)_{b2}\frac{(m_{down})_a}{v}   -\frac{1}{\sqrt{2}} (V^d_L)_{ia} (V^d_L)_bi\frac{(m_{down})_a}{v} \right)\hat d_{a_R}  \nonumber \\
&& + iA\bar{ \hat{u}}_{b_L}\left(  (V^u_L)_{2a} (V^u_L)_{b2}\frac{(m_{up})_a}{v}    -\frac{1}{\sqrt{2}} (V^u_L)_{ia} (V^u_L)_{bi}\frac{(m_{up})_a}{v}\right)\hat u_{a_R}  + \mbox{H.c.}\,,
\label{YcaseIII}
\end{eqnarray}
with the subscript $b=1,2,3$.

Observe that the  interactions in Eqs. (\ref{YcaseI}), (\ref{YcaseII}) and (\ref{YcaseIII}) lead to processes that violate flavor mediated by $A$ as, for example,  $K^0 - \bar K^0$ oscillation. We call the attention to the fact that each variant has its proper set of Yukawa couplings but we used only one set of $V^{u,d}_L$ to diagonalize the different set of mass matrices that appear  in each variant. This is possible because each mixing matrix may diagonalize more than one mass matrix when this mass matrix is Dirac type.

\section{FCNC constraint on $\mathbf{m_A}$ from $K^0 - \bar K^0$ transition}\label{sec:4}
We saw in the previous section that  $A$ mediates processes that violate  flavor at tree level. Those processes may be,  for example,  $K^0(s\bar d) - \bar K^0(\bar s d)$, $D^0(c\bar u)- \bar D^0(\bar c u)$, and  $B^0(d \bar b ) - \bar B^0(\bar d b)$ transitions. For our proposal here, that is to check if FCNC processes are sensitive to the problem of family replication, it is just necessary to investigate the impact of the contribution of $A$  to just one of these transitions because if one is sensitive the other will be, too. However, as we also wishes to obtain a bound on the mass of $A$ we, then, choose to focus on the  $K^0(s\bar d) - \bar K^0(\bar s d)$ transition. This is so because $\Delta m_K$ is much  more restrictive than $\Delta m_D$ and $\Delta m_B$. We, then,  calculate the contributions of $A$ for the three variants discussed in the previous section.  

On requiring that $A$ contribution recovers the experimental results, we  obtain  a lower bound on $m_A$. We will see that such constraints vary for each case  of family discrimination.  For attaining our proposal we must obtain the effective lagrangian responsible for the transition.

From the lagrangian in Eqs.(\ref{YcaseI}), (\ref{YcaseII}), and (\ref{YcaseIII}), following the procedure in Ref. \cite{Gabbiani:1996hi}, we obtain the following effective lagrangian that leads to the $K^0-\bar K^0$ transition,
\begin{eqnarray}
 {\cal L}^{K}_{eff}=-\frac{1}{m^2_A}[\bar d(C^R_K P_R-C^L_K P_L)s ]^2,
 \label{effK}
\end{eqnarray}
where $P_{R,L}$ are the  left-handed and right-handed projections while  the coefficients $ C^{L,R}_K$  are given below.

With this lagrangian in hand, and following the procedure of Ref. \cite{Gabbiani:1996hi},  we obtain the following general expression to the mixing parameter $\Delta m_K$,
\begin{equation}
    \Delta m_K=\frac{2m_K f_K^2}{m^2_A}\left[\frac{5}{24}Re\left((C^L_K)^2+(C^R_K)^2\right)\left(\frac{m_K}{m_s+m_d}\right)^2 + 2Re(C^L_K C^R_K)\left[\frac{1}{24}+\frac{1}{4}\left(\frac{m_K}{m_s+m_d}\right)^2\right] \right]\,,
\end{equation}
where $f_K$ is the decay constant of $K^0$.

As we show below, family discrimination will manifest clearly in the expressions for the coefficients $C^{L,R}$ because each case leads to a specific behavior of $C^{L,R}$ with the elements of $V^{u,d}_L$.

\begin{itemize}
    \item \textbf{Variant I}
    \begin{eqnarray}
   &&  C^R_K= \frac{1}{\sqrt{2}} (V^d_L)_{32}(V^d_L)_{13}\frac{m_s}{v}- (V^d_L)_{i2}(V^d_L)_{1i}\frac{m_s}{v}\,,\nonumber \\ && C^L_K= \frac{1}{\sqrt{2}}(V^d_L)^*_{13}(V^d_L)^*_{32}\frac{m_d}{v}- \frac{1}{\sqrt{2}}(V^d_L)^*_{1i}(V^d_L)^*_{i2}\frac{m_d}{v}\,,
    \label{cofKI}
    \end{eqnarray}
    where $i=1,2$.
    
    \end{itemize}
\begin{itemize}
 \item  \textbf{Varint II}
    \begin{eqnarray}
   && C^R_K= \frac{1}{\sqrt{2}}(V^d_L)_{12}(V^d_L)_{11}\frac{m_s}{v}-\frac{1}{\sqrt{2}}(V^d_L)_{i2}(V^d_L)_{1i}\frac{m_s}{v}\,,\nonumber \\ && C^L_K= \frac{1}{\sqrt{2}}(V^d_L)^*_{11}(V^d_L)^*_{12}\frac{m_d}{v}-\frac{1}{\sqrt{2}}(V^d_L)^*_{1i}(V^d_L)^*_{i2}\frac{m_d}{v}\,,
    \label{cofKII}
    \end{eqnarray}
    where $i=2,3$.
    \end{itemize}
\begin{itemize}
 \item  \textbf{Variant III}
    \begin{eqnarray}
     && C^R_K= \frac{1}{\sqrt{2}}(V^d_L)_{22}(V^d_L)_{12}\frac{m_s}{v}-\frac{1}{\sqrt{2}}(V^d_L)_{i2}(V^d_L)_{1i}\frac{m_s}{v}\,,\nonumber \\ && C^L_K= \frac{1}{\sqrt{2}}(V^d_L)^*_{12}(V^d_L)^*_{22}\frac{m_d}{v}-\frac{1}{\sqrt{2}}(V^d_L)^*_{1i}(V^d_L)^*_{i2}\frac{m_d}{v}\,,
    \label{cofKIII}
    \end{eqnarray}
    where $i=1,3$.
\end{itemize}

Observe that the  process is very sensitive to family discrimination  because $\Delta m_K$ varies  with $\left(V^{u,d}_L \right)^4$.

We now have  in hand all the ingredients necessary for we obtain the contributions of $A$ to the $K^0 -\bar K^0$ meson transition for the three variants and check how sensitive is this transition regarding family discrimination.

\section{Numerical results}\label{sec:5}
The coefficients $C^{L,R}_{K}$ depend on the mixing matrices $V^u_L$ and $V^d_L$ whose entries  are free parameters that obey the constraint $V^{u }_L V^{d \dagger}_L=V_\text{CKM}$. Remember that the idea here is to check if the process $K^0 -\bar K^0$ is sensitive to family discrimination when mediated dominantly by the pseudo-scalar $A$. As usual, we must choose a texture to $V^{u,d}_L$. Here, as  illustrative examples, we use the following textures:

\begin{figure}
    \centering
    \includegraphics[width=\columnwidth]{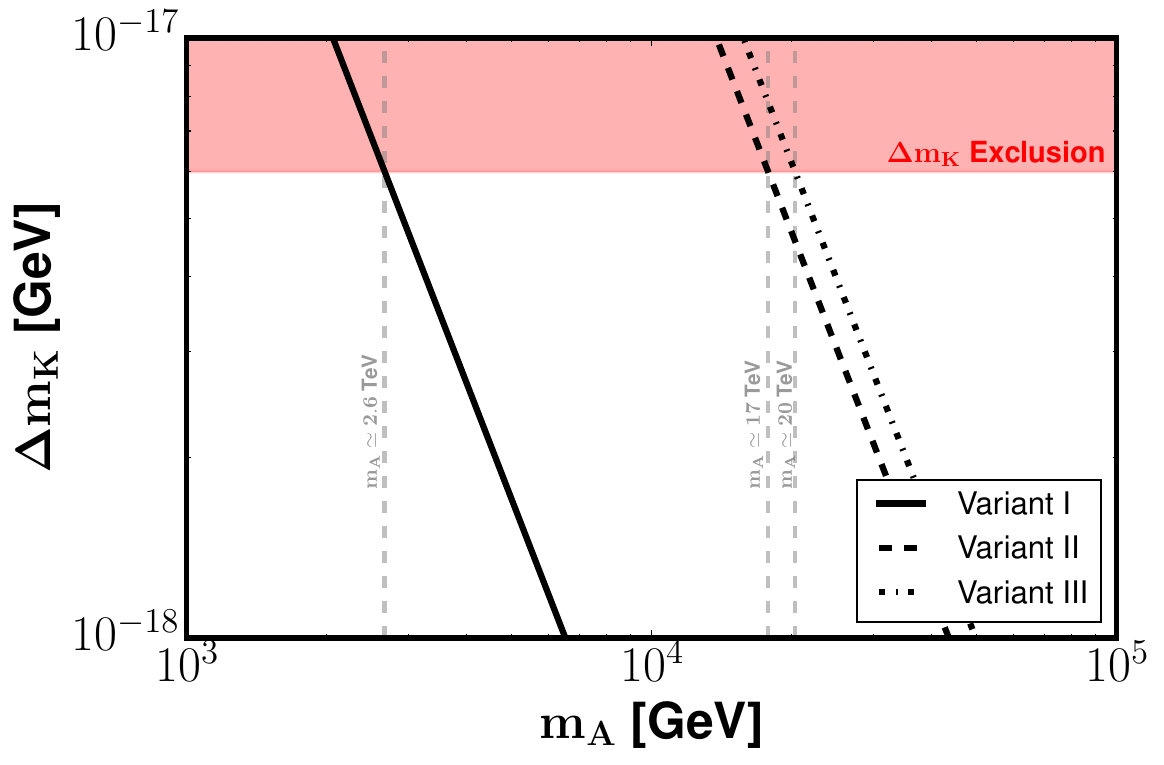}   
    \caption{Numerical results for variant I (continuous curve), variant II (dashed curve) and variant III (dashed dotted curve).  }
  \label{fig:CaseI}
\end{figure}

\begin{equation}
    V^d_L = 
\begin{pmatrix}
 0.849036  &  0.17803  &  0.497437 \\
 0.175894  & -0.983055 &  0.0516103\\
 0.498197  & 0.0436771 & -0.865963  
\end{pmatrix},
\label{EId}
\end{equation}
and
\begin{equation}
    V^u_L = 
\begin{pmatrix}
 0.868672  &  -0.0475631  &  0.493095 \\
 0.38298  & -0.915136 &  0.125881\\
 0.512263  & 0.00476072 & -0.858815  
\end{pmatrix}.
\label{EIu}
\end{equation}

These textures reproduce the experimental values of the elements of CKM matrix and, which is more important, they guarantees that the Higgs contribution to the $K^0(s\bar d) - \bar K^0(\bar s d)$ transition falls inside the experimental error \cite{Oliveira:2022dav}. 

Concerning the VEV's, we take   $v_{\chi^{\prime}}=10^4$ GeV and $v^2_\eta + v^2_\rho=(246\mbox{ GeV})^2$. The other input parameters are found in \autoref{Tab:Input}. These VEV's guarantee that  the mass of $Z^{\prime}$ satisfies the LHC constraint \cite{Coutinho:2013lta,Alves:2022hcp}. 

{In what concern $H_1$, $H_2$ \cite{deSPires:2007wat} and $Z^{\prime}$ \cite{Long:1996rfd} their masses are proportional to $v_{\chi^{\prime}}$ \cite{Long:1996rfd}. By the fact that $m^2_A=\frac{1}{2}f v_{\chi^{\prime}}$, and as $f$ is a free parameter, this pseudoscalar may be the lightest of the 3-3-1 particles that contributes to meson  - antimeson transitions.  This justifies why we are not worried about  the contributions of $Z^{\prime}$, $H_1$ and $H_2$ to $K^0(s\bar d) - \bar K^0(\bar s d)$ transition. 

\begin{table}
\begin{tabular}{ |p{3cm}|p{7cm}|  }
 \hline
 \multicolumn{2}{|c|}{Input Parameters} \\
 \hline
 Quark masses& $K^0$ \\
 \hline
  $m_u= 2.16 \ \text{MeV}$  &   $m_K= 497.611 \ \text{MeV}$ \\
  $m_c= 1.27 \ \text{GeV}$  &   $f_K= 156 \ \text{MeV}$ \\
  $m_s= 93.4 \ \text{MeV}$  &   $\Delta m_K= \left(3.484 \pm 0.006 \right) \times 10^{-12} \ \text{MeV}$ \\
  $m_b= 4.18 \ \text{GeV}$  &   \\
  $m_d = 4.67 \ \text{MeV}$ &  \\
 \hline
\end{tabular}
\caption{Input parameters \cite{Workman:2022ynf}.}
\label{Tab:Input}
\end{table}

The current experimental value for the mass difference $\Delta m_K$ is \cite{HFLAV:2019otj,PDBook,Chen:2021ftn,LHCb:2013zpr}
\begin{eqnarray} \nonumber
    &&\Delta m_K = \left(3.484 \pm 0.006 \right) \times 10^{-12} \text{ MeV}.
\end{eqnarray}
The discrepancy among experimental and theoretical (standard model $(\Delta m)_\text{SM}$) predictions  defines the window to new physics. It happens that the standard model prediction involves a considerable amount of uncertainties due to the QCD corrections \cite{Wang:2019try,Wang:2022lfq}. But even with the uncertainties the standard model prediction is in good accordance with experimental results \cite{Buchalla:1995vs,DiLuzio:2019jyq,DeBruyn:2022zhw}.  In view of this, we adopted the conservative perspective assumed in \cite{CarcamoHernandez:2022fvl}  and  took 
the experimental errors as the window to new physics. In other word, we demand that the 3-3-1 contributions be no larger than the experimental error,
\begin{eqnarray}\label{error}
 &&\Delta m_K= 0.006 \times 10^{-12}\mbox{ MeV}.
\end{eqnarray}
We do this for the contribution of $A$ alone and this is sufficient to derive the most stringent bound on $m_A$ and, as side effect, point out which family of quark is favoured to transform as a triplet by $SU(3)_L$. We  display our findings  in FIG.\ref{fig:CaseI}. According to our illustrative parametrization, case III (the second family transforming as triplet)  suffers the most severe bound \footnote{It is curious because this case  was never developed in literature.}, while case I (third family transforming as triplet) receives the less severe bound. The case II( first family transforming as triplet) receives intermediate bound but just  a little less severe than case III.  Then, our finding indicates that the 331RHN with the third family of quarks transforming as triplet is more energetically favoured, i.e., the CASE I is the case in which the pseudoscalar $A$  gains the smallest mass. This turns  the 331RHN with the third family transforming as a triplet by $\text{SU}(3)_\text{L}$ as  the most attractive case. This is in agreement with previous results in Refs. \cite{Long:1999ij}. 

As we saw above, FCNC is sensitive to family discrimination, but it is the conjunction of FCNC processes with collider phenomenology that will have the power of pointing out which variant is favoured by the model. 

Note that once $m^2_A=\frac{1}{2}f v_{\chi^{\prime}}$, then for the case I, which is the case of interest from now on, we get $f > 1352$ GeV. Our study reinforced the result found in Refs. \cite{Long:1999ij} and is in agreement with the results in Ref. \cite{Cherchiglia:2022zfy} in what concerns the pseudoscalar $A$.

Finally, in the previous cases that considered the contribution of the pseudoscalar $A$ to the $K^0 -\bar K^0$ transition \cite{Cogollo:2012ek,Okada:2016whh} it was considered exclusively the variant in which the third family transforms as a triplet. The difference between these works and our work here is that there they considered the central value of $\Delta m_K$. Here we  followed the approach in Ref. \cite{CarcamoHernandez:2022fvl} which is very conservative and  considers that the contribution be no larger than the experimental error. Second, we considered a parametrization  for $V^{u,d}_L$ that respects the contribution of the neutral scalar that plays the role of the standard Higgs to the $K^0 -\bar K^0$ transitions found in Ref. \cite{Oliveira:2022dav}.

\section{Conclusions}\label{sec:6}
In this work we investigated the sensibility of the $K^0-\bar K^0$ transition, when mediated by the pseudoscalar $A$, to the problem of family  discrimination. In general the main obstacle that we face to  extract any realistic conclusion from  $K^0 -\bar K^0$ transition is the input of the textures of the quark mixing. Our case is not different from other analyses. In order to be as realistic as possible we considered input  textures that  agree with the tiny contribution of the neutral scalar that plays the role of the standard Higgs. Such textures were derived in Ref.\cite{Oliveira:2022dav}. In view of this we showed that the $K^0-\bar K^0$ transition can distinguish the three variants by means of bounds on the mass of $A$. Our results point out that  the third family of quark transforming as a triplet by $\text{SU}(3)_\text{L}$ is the case that supports the lightest pseudoscalar among the three variants. As current lower bound on the mass of the  pseudoscalar in 2HDM lies at TeV region or less\cite{CMS:2019kca}, then the third family transforming as triplet seems to be favoured. 

On the opposite way we can say that FCNC processes as $K^0 -\bar K^0$ transition put the most stringent bounds on the mass of the pseudoscalar of the 331RHN for the three variants of family discrimination with $m_A<2.6$ TeV for the case of the third family transforming as triplet, $m_A<17$ TeV for the case second family transforming as triplet or $m_A<20$ TeV for the case of the first family transforming as triplet. This is the first work studying family discrimination case to case.

Of course we understand that family discrimination, in order to  be deciphered totally,   requires  the  conjunction  of results involving FCNC processes with collider physics.  In this regard, collider physics needs to detect any neutral scalar or gauge bosons that mediate FCNC, and meson transition needs to improve their uncertainties concerning QCD corrections. Until there we must keep expanding and refining our studies  with the aim of understanding as deep as possible  the intriguing  area of flavor physics.

\section*{Acknowledgments}
C.A.S.P  was supported by the CNPq research grants No. 	311936/2021-0 and V.O was supported by CAPES.

\bibliography{bibliography}

\end{document}